\documentclass{article}
\usepackage{graphicx}
\usepackage{cite}
\usepackage{booktabs}
\usepackage{threeparttable}
\usepackage{indentfirst}
\usepackage{url}
\begin{document}
\title{The implementation and optimization of Bitonic sort algorithm based on CUDA }
\author{Qi Mu. Liqing Cui. Yufei Song}
\maketitle

\renewcommand{\abstractname}{ABSTRACT}
\begin{abstract}
\noindent
This paper describes in detail the bitonic sort algorithm,and implements the bitonic sort algorithm based on cuda architecture.At the same time,we conduct two effective optimization of implementation details according to the characteristics of the GPU,which greatly improve the efficiency.
Finally,we survey the optimized Bitonic sort algorithm on the GPU with the speedup of quick sort algorithm on the CPU.Since Quick Sort is not suitable to be implemented in parallel,but it is more efficient than other sorting algorithms on CPU to some extend.Hence,to see the speedup and performance,we compare bitonic sort on GPU with quick Sort on CPU. For a series of 32-bit random integer,the experimental results show that the acceleration of our work is nearly 20 times.When array size is about $2^{16}$,the speedup ratio is even up to $30$.
\end{abstract}

\textbf{KEY WORDS:} Parallel; CUDA; Quick Sort; Bitonic sort

\section{INTRODUCTION}  
\noindent  
Sorting is a well-studied topic and a fundamental problem in computer science. Sorting appears as an internal step of many programs and processes. According to statistics data, it indicates that more than 25 percent of time CPU costs is sorting.Hence, efficient sorting algorithm and implementations are one of the basic keys for performance improvement. Currently, there are many sorting algorithms,Bubble sort,Odd-even sort, Insertion sort,Heap sort,Selection sort,Radix sorting,Quick sort,Bitonic sort...\\
\indent  
There is an increasing need for programming to address parallelism on a variety of architectual approaches. However,due to some physical limitation of technology and material,the frequency level of single-core CPU is also influenced.Today's graphics cards contain very powerful multi-core processors. The processors are specialized for compute-intensive,highly parallel computations. They could be used to assist the CPU in solving problems that can be efficiently data-paralleled. Meanwhile,more and more and more scientists,researchers and software developers are using GPU to accelerate their algorithms and application.\\
\indent
GPU \cite{E.Lindholm.{2008}} processing power and memory bandwidth has obvious advantages in terms of cost and power consumption do not need to pay too much with respect to CPU.Due to the highly parallel graphics rendering, making GPU processing power can be increased by increasing the memory bandwidth and parallel processing unit and a memory unit of the control mode. GPU designers put more transistors used as an execution unit, not like CPU used as a complex control unit and a cache in order to improve the efficiency of a small number of execution units. CPU integer calculation, branch, logic and floating-point arithmetic operations are performed by different units, in addition to a floating point accelerator. Thus, CPU havs different performance to compute tasks of different types. While GPU computes tasks of different types by an integer and floating-point arithmetic unit. so the power of integer GPU computes is similar to its floating-point capability. At present, the mainstream has adopted a unified architecture GPU unit. In adition, GPU computing has a huge advantage, its memory subsystem. The Ultra-high bandwidth of memory not only makes tremendous floating-point capability keep a stable throughput, but guarantees the efficient operation of data-intensive tasks. This is the why GPU is more and more popular in video games, Film industry, industrial design, medical imaging, space exploration, telecommunications, etc.

\section{CUDA}  

\subsection{CUDA Program Model}  
\noindent
CUDA \cite{J. Nickolls.{2008}} is a parallel programming model and software environment for NVIDIA's GPUs. CUDA allows programmer to program kernels executed on the GPU. When one kernel is executed in parallel, a number of CUDA threads are also created at the same time. Threads are organized in warps and blocks. A warp is a fixed-size group of threads (32 threads on GPUs with compute capability 1.x, 2.0), while a block consists of up to 16 (1.x) or 32 (2.0) warps. The number of warps per block and the number of blocks (and therefore the total number of threads) can be specified for each individual kernel launch. When calling kernal, the data transmitted from the CPU main memory to the GPU memory, and is sent back to CPU after calculation process.

\subsection{CUDA Memory Model}  
\noindent
A thread that executes a kernel has access to high-performance thread local registers with the same lifetime as the lifetime of the thread. Each thread block has a shared memory (visible to all threads of the block) with the lifetime of the block. The shared memory is as fast as registers, as long as there are no conflicting accesses by threads of the same block. All threads have random access to global memory that has the lifetime of the application, i.e. data reside in global memory for several kernel launches. In general, global memory is significantly slower than access to registers or shared memory. When threads of the same half-warp (upper or lower 16 threads of a warp) access the global memory simultaneously, the accesses can be coalesced into a single memory transaction, if the accessed memory lies in the same global memory segment. If not, the simultaneous access causes multiple sequential memory transfers. \\
\indent
CUDA provides a barrier mechanism for synchronization of threads of the same block, but does not provide any synchronization mechanisms for threads of different blocks. In general,the host is not blocked when launching a kernel, but it can explicitly wait for a kernel to finish(host synchronization). Since variables reside in global memory for multiple kernel launches, tasks can be synchronized by distributing the workload to multiple kernel launches using host synchronization, while keeping information in global memory.\\
\indent
Efficient synchronization and memory access are two of the most important factors that influence the performance of a GPGPU application. Since kernel launches cause small startup delay, the application should use CUDA 's barrier mechanism instead of host synchronization or even better should avoid explicit synchronization. The host can only access the global memory of the GPU, thus global memory access cannot be completely avoided, but the global memory accesses of a kernel should be reduced to a minimum.
\begin{figure}[htbp]  
\centering
\centerline{\includegraphics[height=5cm,width=5cm]{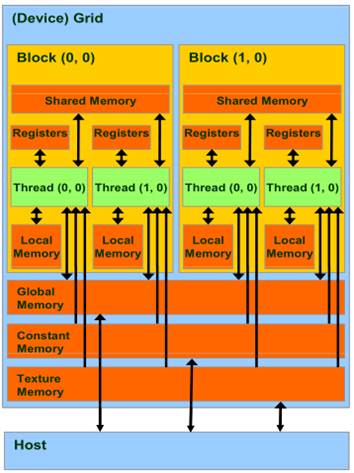}}
\caption{ cuda memory model}
\end{figure}

\section{BITONIC SORT ALGORITHM}  

\subsection{Algorithm Description}
\noindent
Bitonic sort \cite{3} is a binary merge sort, used in the parallel processor with a good parallel performance.Bitonic sequence, for example, {1,5,9,10,12,8,7,2} is a bitonic sequence,the first half of sequence is Monotonically increasing, the second half of this sequence is Monotonically decreasing. Similarly,{12,8,7,2,1,5,9,10}is also a bitonic sequence. Let A be an arbitrary input sequence to sort and let $n=2^k$ be the length of A. The process of sorting A then consists of k phases. The subsequences of length 2\\
\centerline{(A[0],A[1]),(a[2],A[3]),...,A[n-2],A[n-1])}
\\are bitonic sequences by definition. In the first phase these subsequences are sorted using bitonic merge (as shown above) alternating descending and ascending, which makes the subsequences of length 4 bitonic sequences:\\
\centerline{(A[0],A[1],A[2],A[3]),...,(A[n-4],A[n-3],A[n-2],A[n-1])}
\\In the second phase these subsequences of length 4 are sorted alternating descending and ascending, resulting in subsequences of length 8 being bitonic sequences. In the r th phase of bitonic sort the total number of subsequences being sorted is $2^{k-r}$ and the length of each of these subsequences is $2^r$. Sorting a sequence of length $2^r$ using bitonic merge consists of r steps. After the $[k.1]$th phase sequence A is a bitonic sequence. A is sorted in the last phase k. Figure 2 shows the bitonic sorting network for input sequences of size $n=8$. The sorting network consists of $3=log8$ phases, phase p having p steps. Every step consists of $4=n/2$ compare/exchange operations.
\begin{figure}[htbp]
\centering
\centerline{\includegraphics[height=5cm,width=5cm]{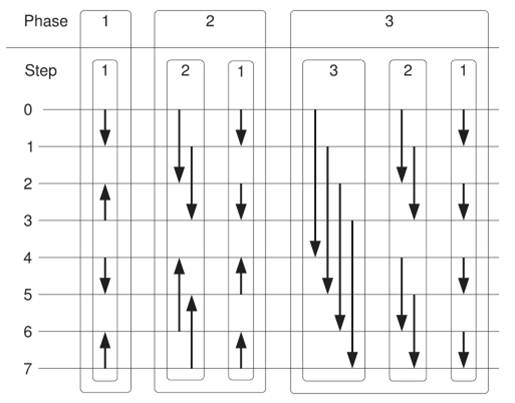}}
\caption{ Bitonic sorting network}
\end{figure}
\noindent

\subsection{Algorithm Analysis} 
\noindent
To our knowledge,quick sort algorithm often used for sorting on the CPU, and high efficiency.Although the average time complexity of quick sort is $O(nlogn)$,it may not be suitable for implementation on the GPU. \\
\indent
Based on the introduction of section 3.1,we know the time complexity of bitonic sort is $O(n(logn)^2)$, it is very suitable for implementation on the GPU for the reasons: the fixed and stable procedure, data independent, and no data cross in each round. It needs $\sum_{i=1}^{logn} i=\frac{logn(logn+1)}{2}$ rounds, at the same time, it completes $\sum_{i=1}^{logn} i*\frac{n}{2}=\frac{nlogn(logn+1)}{4}$ compare-exchange operations.

\subsection{Implementation} 
\noindent
Suppose an array of length $N = 2^k$, the sequence needs $k(k+1)/2$ rounds. And each round calls a kernal function, which completes $N/2$ compare-exchange operations.All these operations can be done by a lot of threads created parallelly, after the end, data is synchronized. Successively,then next kernal is called.

\section{OPTIMIZATIONS}  
\noindent
The method described above has two inevitable shortcomings: too many kernals are called, access global memory consumes too much time, i.e. a long delay.So we consider these two aspects of optimization, reducing the number of memory accesses and the number of kernal launches\cite{Hagen Peters.{2009}}.
\subsection{Optimization1:using shared memory}  
\noindent
In our implementation,every step needs $N/2$ threads,and each thread completes a compare-exchange operation.If a subsequence od length $2^s$ processed in step $s$ completely fits into shared memory of one block,the block first transfers the subsequence into shared memory.That way we can process the steps s,s-1,...,1 using the shared memory and efficient block-synchronization mechanisms between multisteps instead of host-synchronization in one kernal function.So time of accessing global memory is decreased.
\subsection{Optimization2:using the register}  
\noindent
Another optimization is using register.According to the Figure 2,in the phase 3, $4$ address of [0,2,4,6] are used in two successive step 3 and step 2 that we can attach to the register, so that we can reduce two steps to one. Analogously,using this methods can reduce the number of kernal launches and the number of access global memory.

\section{RESULTS}  
\noindent
The test result is obtained on a experimental Platform with CPU(Intel Xeon E5-2620) and GPU(Kepler architecture, K10). Experimental data is 32-bit random integer from 128K to 256M. As shown in below table1.
\begin{table}[htbp]
\begin{threeparttable}[b]
\caption{results}
\centering
\begin{tabular}{crrrrrc}
    \toprule
    &\multicolumn{2}{c}{CPU Times(ms)} & \multicolumn{3}{c}{GPU BitonicSort Times(ms)}\\
    \cmidrule(l{1em}r{1em}){2-3} \cmidrule(l{1em}r{1em}){4-6}
    Array size & QuickSort & BitonicSort & Basic & Semi & Optimized & Ratio\\
    \midrule
    $128K$ & $-$ & $30.00$ & $0.76$ & $0.46$	& $0.36$ &--\\
    $256K$ & $20.00$ & $60.00$ & $1.21$ & $0.87$ & $0.66$ & 30.2\\
    $521K$ & $30.00$ & $110.00$ & $2.22$ & $1.78$ & $1.31$ & 22.7\\
    $1M$ & $80.00$ & $250.00$ & $4.58$ & $3.89$ & $2.80$ & 28.5\\
    $2M$ & $150.00$ & $550.00$ & $8.90$ & $7.95$ & $5.87$ & 25.5\\
    $4M$ & $280.00$ & $1230.00$ & $18.14$ & $16.59$ & $12.30$ & 22.7\\
    $8M$ & $590.00$ & $2670.00$ & $38.13$ & $35.29$ & $26.36$ & 22.3\\
    $16M$ & $1230.00$ & $5880.00$ & $80.09$ & $75.52$ & $56.27$ & 21.8\\
    $32M$ & $2570.00$ & $12900.00$ & $173.77$ & $162.56$ & $120.93$ & 21.3\\
    $64M$ & $5360.00$ & $27780.00$ & $373.52$ & $350.87$ & $258.61$ & 20.7\\
    $128M$ & $11180.00$ & $59860.00$ & $803.16$ & $756.94$	& $553.49$ & 20.1\\
    $256M$ & $23260.00$ & $128660.00$ & $1727.23$ & $1631.92$ & $1185.02$ & 19.6\\
    \bottomrule

\end{tabular}
\begin{tablenotes}
\item [1] Basic : no optimized.
\item [2] Semi  : optimization1.
\item [3] Optimized : optimization1 and optimization2.
\item [4] Ratio : acceleration ratio = Times(CPU Quick Sort)/Times(GPU Bitonic Sort)

\end{tablenotes}
\end{threeparttable}
\end{table}\\
Table 1 shows that quick sort is much faster than bitonic sort on the CPU,this is due to the differences in time complexity. After our two step optimization,the efficiency of bitonic sort algorithm on the GPU has increased significantly.
\section{CONCLUSION AND FUTURE WORK}  
\noindent
We have presented an efficient implementation of bitonic sort based on CUDA.To achieve this we carefully optimized our implementation with respect to the number of accesses to global memory and the number of kernal launches. For a series of 32-bit random integer,our experimental results show that the acceleration of CPU/GPU is more than 20 times.\\
\indent
Due to our current work is a relatively simple, so we will strengthen two main directions in the future work.The first is to test different types of data,such as 64-bit integer,32-bit float,64-bit double.The second is to further explore and compare the performance of a multicore GPU bitonic sort implementation.


\end{document}